\documentstyle[12pt,a4]{article}
\newcommand{\be}{\begin{equation}}
\newcommand{\ee}{\end{equation}}
\newcommand{\bdis}{\begin{displaymath}}
\newcommand{\edis}{\end{displaymath}}

\begin{document}
\centerline{Procedeengs of "III Congresso Nazionale di Cosmologia"}
\centerline{Grado (Ts), 26-28 october 1994}

\centerline{\LARGE Multifractal properties of visible matter distribution:}
\centerline {\LARGE a stochastic model}
\centerline{}
\centerline{}

\centerline{\Large Francesco Sylos Labini$^{(1,2)}$ and
Luciano Pietronero$^{(2)}$}
\date{}
\centerline{}

\centerline{\footnotesize ($^1$) Dipartimento di Fisica, Universit\`a di
Bologna, Italy}

\centerline{\footnotesize ($^2$) Dipartimento di Fisica, Universit\`a di Roma
``La Sapienza''}
\centerline{\footnotesize P.le A. Moro 2, I-00185 Roma, Italy.}

\begin{abstract}
Galaxies and clusters distributions show two
major properties: (i) the positions of galaxies
and clusters are characterized by a power law
distribution indicating properties with respect to their
positions. (ii) The distribution of masses is also characterized
by a power law corresponding to self similarity of
different nature. These two properties are naturally unified
by the concept of multifractality. This concept
naturally arises if the distribution of matter,
as given by both positions and masses, has self similar properties.
We discuss the experimental situation in this
respect and we introduce a simple stochastic
model based of the  aggregation process of particles
The aim of this model is to
understand which characteristic properties of the
aggregation probability
can gives rise to
multifractal distribution
In particular we find that a crucial element in this respect is that
the aggregation probability should depend on
the environment of the aggregation region.

\end{abstract}

\section{Introduction}
The basic characteristic of the observable galaxy distribution
is the existence of
Large Scale Structures (LSS) having fractal distribution
at least up to some length deeper of the available samples:
the two point $\:number-number$ correlation function
is a power law with exponent
$\:\gamma\sim 1.6-1.8$
up to the sample limit for galaxy and cluster distributions
 (Coleman and Pietronero 1992):
\be
G(r) = <n(r)n(0)> \sim r^{-\gamma}
\ee
A second important observational feature is the
galaxy mass function:
this function determines the probability of having a mass in the
range between $\:M$ and $\:M+dM$ per unit volume,
and can be described by the Press-Schechter
function that shows a power law behaviour
followed by an exponential cut-off for very large masses
(Press and Schechter 1974):
\be
n(M)dM \sim M^{\delta-2} exp(-(M/M^{*})^{2\delta})dM
\ee
with $\:\delta \sim 0.2$.
\smallskip

These two observational evidences can be
naturally linked together by the concept
of multifractal (MF) that provides
a unified picture of the mass and spatial distributions.
For an introduction to the concept of MF see Coleman and Pietronero (1992),
and Sylos Labini and Pietronero (1994a).
\smallskip

Masses of different galaxies
can differ by as much as a factor $\:10^{6}$:
and it is important to include these mass values
in order to describe
the entire matter distribution
and not just the galaxy positions.
The distribution of visible matter
is described by the density function:
\be
\rho(\vec{r}) = \sum_{i=1}^{N} m_{i} \delta(\vec{r}- \vec{r_{i}})
\ee
where $\:m_{i}$ is the mass of the $\:i$-th galaxy.
This distribution corresponds to a measure defined on the set of
points which have the correlation properties described by eq.(1).
In this
situation the whole distribution of matter may show self-similar
properties (Pietronero 1987) and this can be described by the more
general concept of multifractal. Therefore the
 MF distribution discussed here is related to the total distribution
of visible matter, including the galaxy masses.
\smallskip

A MF distribution describes systems with
local properties
of self-similarity (Paladin and Vulpiani 1986)
and it is characterised by a continuous set of exponents
{$\:\alpha$} each defined on a set with fractal dimension $\:f(\alpha)$.
\begin{figure}
\vspace{6cm}
\caption {First iterations for the construction
of a self similar distribution that leads to a MF measure.
The support of the
measure is itself a fractal}
\end{figure}
This distribution implies a strong correlation between spatial
and mass distribution (Fig.1) so that the number of objects with
mass $\:M$ in the point $\:\vec{r}$
 per unit volume $\:\nu(M,\vec{r})$, is a function
of space and mass and is not separable
in a space density multiplied a mass function (Binggeli et al. 1988).
It can be shown (Sylos Labini and Pietronero 1994a)
that the mass function of
a MF distribution, in a well defined volume,
has indeed a Press-Schechter behaviour in which the
exponent $\:\delta$ (eq.(2)) can be related to the
properties of  $\:f(\alpha)$.
Moreover the fractal dimension of the support
is $\:D(0)=f(\alpha_{s})=3-\gamma$ (eq.(1)).
 Hence with the knowledge of the
whole $\:f(\alpha)$ spectrum one obtains information on the
correlations in space as well as on the mass function.
\smallskip

Coleman and Pietronero (1992) have performed a MF
analysis of the CfA 1 redshift survey, assigning  to each galaxy a mass
proportional to its luminosity: clearly this  is a crude
approximation, however a better
relation between luminosity and mass
should not change the MF nature of this distribution.
They found that the whole distribution of matter provides unambiguous evidence
for a MF behaviour.
\smallskip

Apart from the determination of the
MF spectrum, there are many  other observational
consequences of the multifractal behaviour
for the galaxy distribution (luminosity
function, number counts, luminosity segregation, etc.),
that will be discussed
in detail in Sylos Labini and Pietronero 1994a.
\smallskip

{}From a theoretical point of view one would like to identify
the dynamical processed that lead to such a MF distribution.
In order
to gain some insight into this complex problem
 we have developed
a simple stochastic model (Sylos Labini and Pietronero 1994b)
that includes the basic properties of
the aggregation process and allows us to
 pose a variety of interesting questions
concerning the possible dynamical origin of the MF distribution.
 The dynamics is characterised by some parameters,
that have a direct physical meaning in term of cosmological processes.
In this way we can relate the input parameters of the dynamics
to the properties of the final configuration and produce a sort of
phase diagram.
The main
point that we want to investigate here, is
whether  the LSS are generated by
an amplification of the small amplitude initial fluctuations or if
the generation of such LSS is intrinsically generated by the
non linear dynamics, that has an
asymptotic critical state. We find the the dependence
from the local environment of the aggregation probability is the crucial
element in order to give rise a fractal (and multifractal) structure.

\section{The model}

In  our model the formation of structures
proceeds by merging
of smaller objects.
When to particles collide in order to form
a bound state, they have to dissipate a certain amount of energy.
The basic physical mechanism responsible of energy dissipation
in a collisionless and pressurless dustlike particles,
interacting only via gravitational force, is the
{\em dynamical friction}: a test particle moving moving
through a cloud of other background particles,
undergoes to a systematic deceleration effect due to the
gravitational scattering (Chandrasekhar 1943).
Clearly this process depends from the local density,
but it depends also from the relative mass ad velocity of
the test particle with respect to the background ones.
Here we consider only the dependence
from the local density: we are currently developing a model
in which we take in account
also of the other parameters of the dynamical friction
(Sylos Labini et al. 1994c).
Due to the effect of the energy dissipation, the aggregation process
dependent on the {\em environment}
in which it takes place, and it is more efficient in more
denser region.
\bigskip

\begin{figure}
\vspace{5cm}
\caption{The probability of making an irreversible aggregation
$\:P_{a}$ during a collision is greater in denser regions
than in sparse ones}
\end{figure}

In our model
when two particles collide they have a probability $\:P_{a}$
of irreversible aggregation and probability $\:1-P_{a}$
to scatter. In such a manner the gravitational interaction is simulated
only through the aggregation probability $\:P_{a}$ that
is made dependent from some parameters,
$\:P_{a} = P_{a}(\alpha,\beta,..)$,
that define the dynamics of the aggregation process.
We find that environment dependence of the dynamical friction
{\em breaks the spatial
symmetry} of the aggregation
process: this is one of the fundamental element
that can give rise to a fractal (and multifractal) distribution.

We have adopted the following procedure  to
estimate the local density,
and introduce the
dissipation effect
in the simulation:
we  assign an influence function to each particle that
describes the contribution of the particle to the dissipation effect.
The aggregation probability is proportional to the
total energy dissipated via dynamical friction.
The influence
function of
the generic
particle in  the point $\:\vec{y}$ with mass $\:m(\vec{y},t)$
at the time $\:t$ on the
point  $\:\vec{x}$, where the collision occurs, is described by:
\be
f(t;\vec{x},\vec{y}) = exp(-\frac{|\vec{y}-\vec{x}|}{m(\vec{y},t)^{\beta}})
\ee
The multiparticle influence function is:
\be
 F(t;\vec{x}) = \int_{V} f(t;\vec{x},\vec{y}) d\vec{y}
\ee
where a suitable value for the volume of integration is chosen.
We define the aggregation probability as:
\be
P_{a}(t,\vec{x}) = \sim  F(t,\vec{x})^{\alpha}
\ee
where we
have introduced two free parameters $\:\alpha$ and $\:\beta$,
that characterized the dynamics,
and we have studied their role in the aggregation process.

\section{The simulations}

We have implemented a simulation in two dimensions
in which the mass is conserved.
At the beginning $\:N$ particles with equal mass (typically $\:N=10^{4}$
 and $\:m=1$)
are distributed randomly over a grid ($\:256^{2}$)  in two dimensions.
To each particle is assigned a velocity of equal module ($\: v=1$)
and random direction. The particles move one step a time along linear
trajectories so that the position of the $\:n$-th particle at time $\:k$ is
defined by:
\be
\vec{x}^{(n)}_{(k)} = \vec{x}^{(n)}_{(k-1)}+ \vec{v}^{(n)}_{(k-1)} \Delta k
\ee
where $\:\Delta k$ is the unitary time step
and $\: \vec{v}^{(n)}_{(k-1)}$
is the velocity of the $\:n$-th particles at the $\:k-1$-th time step.
At each time step the simulation identifies
possible the collision between two o more
particles: typically there are mostly  binary collisions. Once the
collision has been identified
we compute the total influence function due to the particles
around the collision point,
and then the aggregation probability $\:P_{a}$
of forming an irreversible aggregation according to eq.(5).
\smallskip

If the aggregation occurs,
the two particles merge irreversibly in a single particle with
mass
$\:m=m_{1}+m_{2}$.
This aggregate has a probability
$\:P(v)$ of having unitary velocity
ons move along a linear trajectory,
and probability
$\:1-P(v)$ to be stopped.
$\:P(v)$ depends from momentum conservation,
\be
P(v) = \frac{ m_{1} v_{1} + . m_{2}v_{2}}{ m_{1}+m_{2}} .
\ee
Clearly the heavier aggregates stop with a greater probability.
If the aggregation does not occur the two particles will be scattered in
random directions with the same incoming velocities.

To study the spatial properties
of the simulation we compute the integral density-density correlation:
\be
G(\vec{r}) = \int_{0}^{R} d\vec{r} <\rho(\vec{r_{0}})\rho(\vec{r}+\vec{r_{0}})>
 \sim R^{D}
\ee
if $\:D=d=2$ the system is homogeneous while if $\:D<d$ it is fractal.
\smallskip

At the beginning the aggregation is essentially random
and the effect of the local density on the
aggregation process is small everywhere.
Once a certain mass distribution
has been developed, the energy dissipation mechanism becomes dominant
for further aggregation and there is a rapid increment of the aggregation
probability in the are surrounded
by some heavier particles.
\begin{figure}
\vspace{5cm}
\caption{{\em (a): }The integrated density-density correlation function:
the final states has dimension $\:D=1.0$ ($\:\alpha=2.5$
and $\:\beta=0.5$).{\em (b)}: the mass function in the same case}
\end{figure}
In a certain region of the parameter-space for the dynamics,
there is a spatial symmetry breaking of
the aggregation process and
the non linear dynamics generates spontaneously the self-similar fractal
distribution (Fig.3(a)).
The breaking of the spatial symmetry in the aggregation probability
will lead  asymptotically to regions that will never filled,
(voids)  because the density inside is
very low. On the contrary aggregation processes are enhanced in dense region.
The fractal dimension of the final state depends explicitly on the
parameters of the dynamics: $\:D=D(\alpha,\beta)$.
For a larger value of  $\:\alpha$ the aggregation occurs in
only in very dense regions, so that the clustering is  stronger and
the fractal dimension is lower.
The parameter $\:\beta$
tunes the influence of the size of the mass of each particle:
if $\:\beta$ is enhanced the larger aggregate dominates
with respect to the smaller ones.
This mechanism will trigger a spontaneous amplification of fluctuations
and the whole matter distribution will growth in a
in a self-similar way leading to a MF
distribution (Fig.3(b)) (Sylos Labini et al. 1994c).

\section{Conclusion}
We consider an aggregation process
in which the formation of structures is a process that depends on
the {\em local environment}.
This dependence to the energy dissipation
mechanism (dynamical friction)
that strongly depends from the
local density. This environment dependent
aggregation probability breaks spontaneously
the spatial symmetry and leads to the formation of
complex structures.

We have implemented a simulation in two dimension
in which the mass is conserved and the particles move along linear
trajectories.
This model shows an asymptotic fractal distribution
The non linear dynamics leads spontaneously the
self-similar (multifractal) fluctuations
of the asymptotic state, so
that there is not any crucial dependence from initial conditions.
The fractal dimension of the asymptotic state depends only on
the parameters of the non linear dynamics.
The necessary ingredients for a dynamics in order to generate a fractal
(multifractal including masses) distribution
are the breaking of the spatial symmetry,
and the Self-Organised nature of the dynamical mechanism.

\section*{Aknowlegements}
We thank R. Capuzzo-Dolcetta for useful discussions
about the dynamical friction and for computer
facilities.

\section*{References}
- Bingelli,B., Sandage,A., Tammann,G.A. 1988 Ann.
 Rev. Astron. \& Astrophys. , 26 509 \\
- Chandrasekhar, S. 1943 Rev.Mod.Phys. 15,1 \\
- Coleman, P.H. \& Pietronero, L.,1992 Phys.Rep. 231,311 \\
- Chandrasekhar, S. 1942 Principles of stellar dynamics, Dover New, \\
- Paladin,G., Vulpiani,A., 1987 Phys. rep. 156,147 \\
- Pietronero,L., 1987, Physica A 144,257 \\
- Press, W.H, Schechter, P., 1974 ApJ 187,425 \\
- Sylos Labini, F., \& Pietronero, L. 1994a in pre-print \\
- Sylos Labini, F., \& Pietronero, L. 1994b, in "Birth of
the Universe and fundamental physics", F. Occhionero editor,
Springer Verlag, in press \\
- Sylos Labini F. et al., 1994c unpublished
\end{document}